\documentclass[prl,twocolumn,secnumarabic,amssymb, amsmath, nobibnotes, showpacs, aps]{revtex4}
\usepackage{bm}%
\usepackage{epsfig}%
\usepackage{graphicx}%
\expandafter\ifx\csname package@font\endcsname\relax\else
 \expandafter\expandafter
 \expandafter\usepackage
 \expandafter\expandafter
 \expandafter{\csname package@font\endcsname}%
\fi

\begin{document}

\title{Spin-rotation coupling in non-exponential decay of hydrogenlike heavy ions}%

\author{G. Lambiase$^{a,b}$, G. Papini$^{c,d,e}$, G. Scarpetta$^{a,b,e}$}
\affiliation{$^a$Dipartimento di Fisica "E.R. Caianiello"
 Universit\'a di Salerno, 84081 Baronissi (Sa), Italy.}
  \affiliation{$^b$INFN, Sezione di Napoli Italy.}
  \affiliation{$^c$Department of Physics, University of Regina, Regina, SK, S4S 0A2, Canada.}
  \affiliation{$^d$Prairie Particle Physics Institute, Regina, SK, S4S
  0A2, Canada}
  \affiliation{$^e$International Institute for Advanced Scientific Studies, 89019 Vietri sul Mare (SA), Italy.}
\def\be{\begin{equation}}
\def\ee{\end{equation}}
\def\al{\alpha}
\def\bea{\begin{eqnarray}}
\def\eea{\end{eqnarray}}

\begin{abstract}

We discuss a model in which a recently reported modulation in the
decay of the hydrogenlike ions ${}^{140}$Pr$^{\, 58 +}$ and
${}^{142}$Pm$^{\, 60 +}$ arises from the coupling of rotation to the
spin of electron and nuclei (Thomas precession). A similar model
describes the electron modulation in muon $ g-2$ experiments
correctly. Agreement with the GSI experimental results is obtained
for the current QED-values of the bound electron g-factors,
$g({}^{140}$Pr$^{\, 58 +})=1.872$ and $g({}^{142}$Pm$^{\, 60
+})=1.864$, if the Lorentz factor of the bound electron is $\sim
1.88$. The latter is fixed by either of the two sets of experimental
data. The model predicts that the modulation is not observable if
the motion of the ions is linear, or if the ions are stopped in a
target.
\end{abstract}

\pacs{23.40.-s, 27.60.+j}

\maketitle

Recent experiments carried out at the storage ring ESR of GSI in
Darmstadt \cite{GSI,geissel} reveal an oscillation in the orbital
electron capture and subsequent decay of hydrogen-like
${}^{140}$Pr$^{\, 58 +}$ and ${}^{142}$Pm$^{\, 60 +}$. The
modulation has a period of $7.069(8)\,s$  and $7.10(22)\,s$
respectively in the laboratory frame and is superimposed on the
expected exponential decay. The "zero hypothesis" of a pure
exponential decay has been excluded at $99\%$ C.L. for both ions.
Equally improbable causes seem to be periodic instabilities in the
storage ring and detection apparatus. The effect has been
tentatively attributed to neutrino flavor mixing, but this
interpretation presents difficulties that have been discussed in
the literature \cite{mixing}.

It may be useful to analyze the effect of spin-rotation coupling
on the ESR modulation effect. It is in fact the spin part of the
Thomas precession, or spin-rotation coupling, that is responsible
for a similar modulation in $g-2$ experiments. The corresponding
modulation in the detection of electrons produced in the in-flight
decay of muons in storage rings yields the anomalous part of the
muon magnetic moment directly \cite{kinoshita}.

An important point raised in work on ${}^{140}$Pr$^{\, 58 +}$ and
${}^{142}$Pm$^{\, 60 +}$ is that in the initial state these ions,
with nuclear spin $I=1$, and the bound electron can have total
angular momentum $1/2$ when the electrons and nucleus spins are
antiparallel, or $3/2$ when the spins are parallel. The final
state, however, can only have spin 1/2 because decay from the spin
$3/2$ state is forbidden by the conservation of the $F$ quantum
number.

For the sake of simplicity, we treat the nuclei as spin-$1$
particles interacting with their single electron by an amount
sufficient to drag it along the same cyclotron orbits in the
storage ring. We use units $ \hbar=c=1$, but re-introduce standard
units when discussing the results.

Following Bell and Leinaas \cite{bell}, we write the full
Hamiltonian that describes the behavior of nucleus and bound
electron in the external induction field ${\bf B}$ of the ring as
 \[
 H=H_0+H^{(e)}+H^{(n)}\,,
 \]
where
 \[
H_0=H_0^{(e)}+H_0^{(n)}\,,
 \]
\begin{eqnarray}
  H^{(e)} &=& - 2 g_e\mu_B{\bf s}\cdot {\bf B}-{\bm
  \omega}_e \cdot \left({\bf L}+{\bm s}\right)\,,
  \label{He}\\
  H^{(n)} &=& -g_n\mu_N{\bf I}\cdot {\bf B}-{\bm
  \omega}_n\cdot \left({\bf L}+{\bf I}\right)\,. \label{Hi}
\end{eqnarray}
$H_0^{(e, n)}$ contain all the usual standard terms (Coulomb
potential, spin-orbit coupling, spin-spin coupling, etc.), $g_e$
and $g_n$ are the $g$-factors, and $\mu_B$ and $\mu_N$ are the
Bohr and nuclear magnetons.

The Hamiltonians $H^{(e)}$ and $H^{(n)}$ contain the Pauli spin
matrices via $ {\bm s}={\bm \sigma}/2$ and the $3\times 3$ matrices
${\bf I}$ and the last terms account for the Thomas precession. If
the orbits in the ring are stable, the coupling of ${\bm \omega}$ to
the orbital angular momentum ${\bf L}$ may be neglected \cite{bell}.
We will also neglect, for simplicity, any stray electric fields and
electric fields needed to stabilize the orbits and all quantities in
(\ref{He}) and (\ref{Hi}) will be taken to be time-independent.

One can also arrive at (\ref{He}) and (\ref{Hi}) starting from the
Dirac equation \cite{bell, hehl,pap,singh} and the corresponding
relativistic wave equation for spin-1 particles \cite{pap}. The
importance of spin-rotation coupling has also been greatly
emphasized, in a different context, by Mashhoon \cite{mash}.

$H_0^{(e)}$ and $H_0^{(n)}$ contribute to the overall energy
$E^{(e)}$ and $E^{(n)}$ of the states. By referring (\ref{He}) and
(\ref{Hi}) to a left-handed tern of axes rotating about the
$x_2$-axis in the clockwise direction of the ions and with the
$x_3$-axis tangent to the ion orbit in the direction of its
momentum \cite{bell}, we find that the operators governing the
evolution of the spin states are
\begin{equation}\label{Me}
M^e= \left(\begin{array}{cc} E^e &
            \displaystyle{i\left(\frac{\omega_e}{2}-\mu_e B\right)}\\
                \displaystyle{-i\left(\frac{\omega_e}{2}-\mu_e B\right)} &
                E^e  \end{array}\right)
 \end{equation}
and
\begin{equation}\label{Mi0}
M^n_0= \left(\begin{array}{ccc}
E^n-i\displaystyle{\frac{\Gamma}{2}}&
            \displaystyle{i\left(\omega_n-\mu_n B\right)} & 0\\
                \displaystyle{-i\left(\omega_n-\mu_n B\right)} &
                E^n-i\displaystyle{\frac{\Gamma}{2}} &0 \\
                0 & 0 & E^n-i\displaystyle{\frac{\Gamma}{2}}
                \end{array}\right)\,,
\end{equation}
where $\Gamma$ is the energy width of the nucleus. Instead of
(\ref{Mi0}), we consider the sub-matrix
\begin{equation}\label{Mi} M^n=
\left(\begin{array}{cc} E^n-i\displaystyle{\frac{\Gamma}{2}}&
            \displaystyle{i\left(\omega_n-\mu_n B\right)}\\
                \displaystyle{-i\left(\omega_n-\mu_n B\right)} &
                E^n-i\displaystyle{\frac{\Gamma}{2}}
                \end{array}\right)\,,
\end{equation}
because only the non-zero, off-diagonal terms are relevant.

Before the ion decays, the nucleus and the electron separately can
be represented by \cite{papini}
\begin{equation}\label{state}
  |\psi^l(t)\rangle = a^l(t)|\psi^l_+ \rangle + b^l(t)|\psi^l_-
  \rangle\,,
\end{equation}
where $l=e, n$ and $|\psi^l_+ \rangle$ and $|\psi^l_- \rangle$ are
the two helicity states. The coefficients $a^l(t)$ and $b^l(t)$
evolve in time according to the equation
\begin{equation}\label{stateevol}
  i\frac{\partial}{\partial t} \begin{pmatrix}
  a^l(t) \cr b^l(t)\,\end{pmatrix}= M^l \begin{pmatrix}
  a^l(t) \cr b^l(t)\,\end{pmatrix}\,,
\end{equation}
which can be solved by diagonalizing the matrices $M^l$.

The spin flip probabilities for electron and nucleus are
\begin{equation}\label{Pe}
  P^{(e)}=|\langle \psi^e_+|\psi^e(t)
  \rangle|^2=\frac{1}{2}\left[1-\cos\Omega_e t\right]
\end{equation}
and
\begin{equation}\label{Pi}
  P^{(n)}=\frac{e^{-\Gamma t}}{2}\left[1-\cos\Omega_n t\right]\,,
\end{equation}
where
 \[
 \Omega_e= g_e \mu_B B-\omega_e\,, \quad \Omega_n=2(\mu_n B-\omega_n)\,.
 \]
The probability for electron and nucleus to flip their spins at
time $t$, if at $ t=0$ their spins were both antiparallel to the
nucleus momentum, is
\begin{equation}\label{PePi}
  P\propto P^{(e)}P^{(n)}=\frac{e^{-\Gamma t}}{4}\Big\{
  1-\cos\Omega_n t-\cos \Omega_e t+
\end{equation}
 \[
 +\frac{1}{2}\Big[ \cos(\Omega_e+\Omega_n)t+ \cos(\Omega_e-\Omega_n)t
 \Big] \Big\}\,.
 \]
We use the uncorrelated probability (\ref{PePi}) because in our
model the electron and nucleus spins are treated as independent.
The second and third terms in (\ref{PePi}) represent the spin-flip
angular frequencies of nucleus and electron. The latter frequency,
when applied to the muon, gives the muon anomaly in $g-2$
experiments. The fourth term represents transitions to the
forbidden spin-$3/2$ final state with angular frequency $
\Omega'\equiv \Omega_e + \Omega_n$. The last term yields the
probability that the electron and nucleus spins are antiparallel,
which is the probability of interest. We obtain
\begin{equation}\label{Pfinal}
  P'\simeq \frac{e^{-\Gamma
  t}}{4}\Big[1+\frac{1}{2}\cos(\Omega_e - \Omega_n)t \Big]\,.
\end{equation}
The angular frequency $\Omega \equiv\Omega_e - \Omega_n$ with which
the electron and ion spins find themselves antiparallel and which,
therefore, enables the ion decay, must now be related to laboratory
system quantities.

We now follow \cite{bell,jackson,jacksonbook} and re-introduce
standard units.

The calculation of the precession frequency of the electron spin in
its motion about the heavy nucleus first takes into account the
contribution given by the first term in (\ref{He}) which in the
nucleus rest frame is given by ${\bm \omega}_g =
-\displaystyle{\frac{eg}{2m}{\bf B}}$. We also assume that $\langle
{\bm \beta}_e \cdot {\bf B}_e \rangle =0 = \langle {\bm \beta}_e
\times {\bf E}_e \rangle$, where ${\bf B}_e$ and ${\bf E}_e$ are the
magnetic and electric fields in the nucleus rest frame and ${\bm
\beta}_e= {\bf v}_e/c$ is the velocity of the electron relative to
the nucleus. The spin-rotation term in (\ref{He}) is given, in the
rest frame of the electron, by ${\bm \omega}_{T\,, rf} =
\displaystyle{\frac{\gamma_e}{\gamma_e+1}}{\dot {\bm
\beta}_e}\times{\bm \beta}_e$, where ${\dot {\bm \beta}_e}$ is the
electron acceleration, and $\gamma_e=1/\sqrt{1-\beta_e^2}$ is the
Lorentz factor. ${\bm \omega}_{T\,, rf}$, hence ${\dot {\bm
\beta}_e}$, are generated by the Coulomb force ${\bf F}_C=e{\bf
E}_n$, where ${\bf E}_n$ is the electric field of the charged
nucleus, and by the Lorentz force ${\bf F}_L=e({\bf E}^\prime+{\bm
\beta}_e\times {\bf B}^\prime$), where ${\bf E}^\prime$ and ${\bf
B}^\prime$ are the electric and magnetic fields in the rest frame of
the electron. They are related to the magnetic field of the storage
ring by means of a Lorentz transformation. Since the electron
propagates in an external magnetic field, its velocity is shifted by
$\sim \mu_B B |\psi_{100}|^2$. This term, however, turns out to be
negligible.

The electric field ${\bf E}_n$ gives rise to the anomalous Zeeman
effect. This term is not relevant to our study and will not be
considered further.

The force ${\bf F}_L$, on the other hand, generates an
acceleration of the bound electron that contributes to the Thomas
precession, hence to the precession of the electron spin with
respect to the magnetic field ${\bf B}$. Passing from the electron
rest frame to the nucleus frame and using the Lorentz
transformation of the electric and magnetic fields, the Thomas
precession becomes ${\bm \omega}_T =
\displaystyle{\frac{e}{m_e}\frac{2(\gamma_e-1)}{\gamma_e}{\bf
B}}$. Therefore, the precession frequency of the electron is given
by $\displaystyle{\frac{d{\bf s}}{dt}}|_{\mbox{nf}}={\bm \Omega}_e
\times {\bf s}$, where
\begin{eqnarray}
  {\bm \Omega}_e & =&{\bm \omega}_g + {\bm \omega}_T = -\frac{2\mu_B}{\hbar}
  \left(a_e-1+\frac{2}{\gamma_e}\right){\bf B} \nonumber \\
   & \simeq & - 1.7587 \times 10^{11}
   \left(a_e-1+\frac{2}{\gamma_e}\right)\frac{B}{\text T}\mbox{Hz}\,\,{\bf u}_2\,. \label{Omega-e}
\end{eqnarray}
$a_e=(|g_e|-2)/2$ is the electron magnetic moment anomaly, and ${\bf
u}_2$ is a unitary vector direct along the $x_2$-axis. It is
parallel to the magnetic field of the storage ring and therefore
orthogonal to the circular orbit of the nucleus.

If the nucleus moves along a circular orbit in the presence of only
the external magnetic field, then the precession frequency of the
nucleus spin, in the reference frame comoving with the nucleus, is
$\displaystyle{\frac{d{\bf I}}{dt}}|_{\mbox{nf}}={\bm \Omega}_n
\times {\bf I}$ \cite{jackson,jacksonbook}, where
\begin{eqnarray}
  {\bm \Omega}_n & = &
  2\left(\frac{\mu_n}{\hbar}\frac{M}{Q}-1\right)\frac{Z}{A}\frac{e}{m_p}{\bf
  B} \simeq 2\left(\frac{{\tilde \mu}A}{2Z}-1\right)\frac{Z}{A}\frac{e}{m_p}{\bf
  B}  \nonumber \\
  & \simeq & 4.785 \times 10^{7} \frac{B}{\text T}\mbox{Hz}\,\,{\bf u}_2\,.  \label{Omega-i}
\end{eqnarray}
$A$ and $Z$ are the mass and charge numbers, and we have written
$\mu_n={\tilde \mu}(e\hbar/2m_p)$. The spin of the nucleus does
not precess if $ \mu_{n}M/\hbar Q =1$. In (\ref{Omega-i}) we have
approximated $ M$ by $ A m_p$. The calculation of $ M$ can be
improved by including contributions from nucleon binding energies,
nuclear surface and Coulomb repulsion effects and the Pauli
exclusion principle. Not all these contributions are positive.

We must now calculate the relative spin precession that the motions
of electron and nucleus generate. From (\ref{Omega-e}),
(\ref{Omega-i}) we obtain
\begin{equation}\label{Omega}
 \Omega=-\left[a_e -1 +\frac{2}{\gamma_e}+\left(\tilde{\mu}-\frac{2Z}{A}\right)
 \frac{m_e}{m_p}\right]\frac{eB}{m_e}\,.
\end{equation}
The magnetic field is related to the the angular velocity of the
particles in the laboratory frame $2\pi f$ by
 \[
 \frac{QB}{M}=\frac{\gamma^2(2\pi f)^2 \rho}{\beta c}\,,
 \]
where $\gamma=1/\sqrt{1-\beta^2}$ is the Lorentz factor, and $\rho$
is the radius of the orbits. The revolution frequency $f$ of the
primary beam can be obtained from $f=v/L$, where $L$ is the length
of the closed orbit, and from the definition of magnetic rigidity
$B\rho=Mv\gamma/Q$ \cite{geissel}. We get
\begin{equation}\label{f}
  f=\frac{e}{m_p}\frac{Z}{A}\frac{B\rho}{L}\frac{1}{\sqrt{1+\displaystyle{\left(
  \frac{Ze}{Am_p}\frac{B\rho}{c}\right)^2}}}\simeq 2 \mbox{MHz} \,.
\end{equation}

It is the frequency $\Omega/2\pi$ that must be compared with the
experimental signal $0.14\, Hz$ found for ${}^{140}$Pr$^{\, 58 +}$
and ${}^{142}$Pm$^{\, 60 +}$ by means of the equation
\begin{equation}\label{Omega-f}
  \Omega= 2 \pi\, 0.14 \, \mbox{Hz}\,.
\end{equation}
The calculation of $g$-factors, based on bound state (BS) QED, can
be carried out with accuracy even though, in our case, the
expansion parameter is $Z\alpha\simeq 0.4$. The results agree with
available direct measurements \cite{vogel}-\cite{moskov}. In
particular, the BS-QED calculation given in \cite{blundell}
includes radiative corrections of order $ \alpha/\pi$ and exact
binding corrections. It yields
\begin{equation}\label{gb}
g_e^b =2\left[\frac{1+2\sqrt{1-(\alpha
Z)^2}}{3}+\frac{\alpha}{\pi}C^{(2)}(\alpha Z)\right]\,,
\end{equation}
where $C^{(2)}$ can be approximated by
\[C^{(2)}(\alpha Z)\simeq \frac{1}{2}+\frac{1}{12}(\alpha Z)^2
+\frac{7}{2}(\alpha Z)^4\,.
\]
The values of $g_e^b $ calculated by applying (\ref{gb}) to
$C^{5+}$ and $O^{7+}$ are in good agreement with the experimental
results. The same formula (\ref{gb}) extrapolated to $ Z= 59$ and
$Z=61$, gives $g_e^b \simeq 1.87205$ for ${}^{140}$Pr$^{\, 58 +}$
and $g_e^b \simeq 1.86365$ for ${}^{142}$Pm$^{\, 60 +}$. The
addition of more expansion terms following the formulae given in
\cite{CODATA} does not change these results appreciably. The
corresponding anomalous parts become respectively $ a_e \simeq
-0.06397$ and $ a_e \simeq -0.06817$. The only {\it free}
parameter in (\ref{Omega}) is the electron Lorentz factor
${\gamma}_e$. It can be fixed by using one set of experimental
data and using the result in the second set. Alternatively, we can
choose as free parameter the {\it average} distance $R$ of the
electron from the nucleus and estimate the electron velocity by
means of the equation $ m_e \gamma_e v_e^2
=\displaystyle{\frac{1}{4 \pi \varepsilon_0}\frac{Ze^2}{R}}$. We
obtain $ \beta_e
=\sqrt{\displaystyle{\frac{b}{2}\left(\sqrt{4+b^2}-b\right)}}$,
where $b \equiv \displaystyle{\frac{1}{4
\pi\varepsilon_0}\frac{Ze^2}{m_e c^2 R}}$. From the experimental
values $L=108.3\,$m, $B\rho=6.44\,$Tm, $\gamma=1.43$,
$\beta=0.71$, assuming that ${\tilde \mu}=2.5$ as given in
\cite{GSI}, using the values of $a_e$ just found and treating $R$
as a parameter, we find that equation (\ref{Omega-f}) has the
solutions $\gamma_e(\text{Pr}^{58+}) \sim 1.88135$ and
$\gamma_e(\text{Pm}^{60+}) \sim 1.87392$, which correspond to $R
\simeq 123\, \text{fm} \simeq 0.145\, a_Z$ for Pr$^{58+}$, and $ R
\simeq 128 \,\text{fm}\sim 0.151 \, a_Z$ for Pm$^{60+}$. Here
$a_Z(=a_0/Z)$ is the Bohr radius ($a_0\sim 0.53 \times
10^{-10}$m). As a comparison, the nuclear radius is given by $R_n
\sim  R_0 A^{1/3}\,\text{fm} \sim 6.2\, \text{fm}$, in nuclear
models in which the nucleons are uniformly distributed in the
nucleus ($R_0\sim 1.2\, $fm).

In our calculations we have adhered to the assignment ${\tilde
\mu}=2.5$ \cite{GSI}. However, a need for a re-measurement of
nuclear magnetic moments arises from improved ways to probe QED
effects, as pointed out in \cite{gustavsson}.

Eq. (\ref{Omega}) does not depend on $ \hbar$. This should be
expected because the electron orbits and those to which (\ref{f})
refers are classical. This is also the extent to which the
treatments of particles in storage rings given in \cite{bell} and
\cite{jackson} agree (see the discussion given in \cite{bell}). Bell
and Leinaas consider quantum fluctuations away from the classical
orbit. We neglect them, in agreement with Jackson \cite{jackson}.

The model has consequences that must be emphasized.

Identical ions in rings of different radii $\rho_1$ and $\rho_2$,
but with the same magnetic rigidity, and the same speed, give rise
to $\Omega\sim \rho^{-1}$. Therefore, the effect becomes smaller in
larger rings for fixed $A/Z$, $g_e$ and ${\tilde \mu}$.

It is also clear that according to our hypothesis, which is
essentially based on the coupling between rotation and the spins of
electron and nucleus, the GSI effect disappears when $f=0$.
Accordingly, the effect {\it can not be measured in experiments
based on the linear motion of the ions, nor can it be observed by
stopping ions in thick aluminum foils} \cite{vetter}.

A more complete model would also require the relativistic
treatment of hydrogen-like atoms (with large nuclear charge) in a
noninertial reference frame.

In summary, the GSI experiments measure $ \Omega$ in our model and
are well suited to study nuclear radii, $g_e$ and ${\tilde
\mu}(g_e)$. This is important in fields like nuclear physics, QED,
BS-QED and stellar nucleosynthesis \cite{takahashi}.

\vspace{0.2in} G.L. and G.S. acknowledge the financial support of
MIUR through PRIN 2006 Prot. $1006023491_{-}003$, of a contract
with the Agenzia Spaziale Italiana, and of research funds provided
by the University of Salerno. G.P. thanks the Department of
Physics of the University of Salerno for kind hospitality.

\end{document}